# An observational test of common-envelope evolution

M. J. Sarna[1], V. S. Dhillon[2], T. R. Marsh[3] and P. B. Marks[4]

[1] *N. Copernicus Astronomical Centre, Polish Academy of Sciences, ul. Bartycka 18, 00–716 Warsaw, Poland.*
[2] *Royal Greenwich Observatory, Apartado de Correos 321, Santa Cruz de La Palma, 38780 Tenerife, Canary Islands, Spain.*
[3] *Astrophysics, Physics Department, The University, Southampton SO9 5NH, UK.*
[4] *Astronomy Centre, School of Mathematical and Physical Sciences, University of Sussex, Falmer, Brighton BN1 9QH, UK.*

**ABSTRACT**
By analysing and modelling the change in the abundance ratio of $^{12}C/^{13}C$ and $^{16}O/^{17}O$ on the surface of the lower mass star of a binary during the common-envelope (CE) phase of evolution, we propose a simple observational test of the CE scenario. The test is based on the infrared measurement of either the $^{12}C/^{13}C$ or $^{16}O/^{17}O$ ratio of red dwarfs in post-common envelope binaries (PCEB's). In certain cases (main-sequence red dwarf secondaries in PCEB's without planetary nebulae), as well as determining whether or not accretion has occurred during the CE phase, we can determine the amount of mass accreted during the CE phase and hence the initial mass of the red dwarf component prior to the CE phase. In the other cases considered (low-mass red dwarfs in PCEB's and red dwarf's in PCEB's with planetary nebulae) we can only say whether or not accretion has occurred during the CE phase.

**Key words:** accretion, accretion discs – nuclear reactions, nucleosynthesis, abundances – infrared: stars – binaries: spectroscopic – stars: abundances – stars: novae, cataclysmic variables

## 1 INTRODUCTION

The most widely accepted scenario for the evolution of cataclysmic variables (CVs) is the common envelope model of Paczyński (1976). According to this model, the progenitors are two main–sequence stars of unequal mass with an initial orbital period of months to years. The zero-age orbital parameters of the system are such that the more massive (primary) star fills its Roche lobe when it reaches its giant or supergiant phase, while its lower mass (secondary) companion remains on the main sequence. Under these conditions, mass transfer to the secondary is dynamically unstable and occurs at such a high rate that the transferred material cannot be incorporated by the secondary and forms a common envelope (CE) around the inner binary. Through the action of friction, the main-sequence star spirals towards the core of the giant, generating luminosity which drives off the CE. What remains is often called a pre-CV or, more generally, a post-common envelope binary (PCEB), typically consisting of a white dwarf and detached red dwarf star with an orbital period of order days (eg. Catalán et al. 1994ab). These systems are thought to become CVs when magnetic braking or gravitational radiation extract sufficient orbital angular momentum for the red dwarf to fill its Roche lobe (see King 1988 and references therein).

Although there is general agreement that the above model applies to CV evolution, there is little direct evidence to support it other than the fact that we observe stars at the various stages of evolution outlined above (see Trimble 1993 and references therein). It is the purpose of this paper to present a direct test of the theory which can be easily confirmed by observations.

## 2 THE MODEL

It is a well known fact that the ratios of various isotopes change during different phases of stellar evolution. The most commonly studied ratios are probably those of $^{12}C/^{13}C$ and $^{16}O/^{17}O$. In a typical main-sequence star, such as the Sun, these ratios take the values 90 and 2724, respectively (Trimble 1993). During main-sequence evolution the distribution of the CNO elements in the hydrogen-burning core changes. These changes occur rapidly when the CNO tri-cycle is dom-inant, ie. for stars of mass greater than 1.4 M$_\odot$, but in lower mass stars these reactions are less efficient. The central abundance of carbon decreases from cosmic to an equilibrium value equal to about 5 per cent of the initial one. At the same time, nitrogen increases, but the oxygen in the star remains at approximately cosmic abundance. The region within which the carbon is being reduced with respect to nitrogen expands outwards with time to greater mass fractions. The carbon distribution becomes essentially fixed at the moment when the central hydrogen content has dropped by about 15 per cent; this is caused by envelope expansion and a temperature decrease in the region where the CNO tri-cycle works. At that time, in the inner region we have roughly equal abundances of $^{12}$C and $^{13}$C, while in the outer region the $^{12}$C/$^{13}$C ratio is approximately solar ($\sim$90). The ratio of $^{16}$O/$^{17}$O in the outer region (75 per cent of the total mass of the star) is constant ($\sim$2724), while in the inner region this ratio is decreased by one order of magnitude. This situation exists up to the moment when the star evolves to become a red giant or supergiant and a thick convective envelope develops. The $^{12}$C/$^{13}$C and $^{16}$O/$^{17}$O ratios change drastically due to convective mixing of the inner and outer regions of the star, becoming 13 and 370 respectively (Harris, Lambert and Smith 1988; El Eid 1994 and references therein).

It is probable that the dwarf star accretes material during the CE phase. Since, during the CE phase, the red-dwarf secondary effectively exists within the atmosphere/envelope of the giant or supergiant primary, the accreted material has the abundances/composition of a giant/supergiant star. The important question is how much material is accreted? Iben and Livio (1993) recently modelled this accretion and found a wide range in the accreted masses, this range depending on the evolutionary scenario and the component masses adopted. Livio & Soker (1984) modelled the evolution of star-planet systems through CE phase, they found that there exists a critical planetary mass below which the planet is totally evaporated, and above which the planet is transformed into a low mass stellar companion to the giants core. Therefore, from a theoretical point of view, the amount of mass accreted is still open to question. However, we have restricted our models to two evolutionary scenarios and constrained the component masses from observations.

We know that many PCEBs have red dwarf masses in the range 0.1 to 0.4 M$_\odot$ and typical white dwarf masses in the range 0.5 to 0.65 M$_\odot$ (de Kool & Ritter 1993; Catalán – private communication). However there are systems where the red dwarf has a mass lying outside the range given above, eg. V471 Tau and AA Dor. The range of white dwarf masses is in agreement with a recent study of the central stars of planetary nebulae (Stasińska & Tylenda 1990) where the large majority of remnant-star masses are found to be in the range 0.55 to 0.65 M$_\odot$. Two evolutionary scenarios and hence two types of progenitor pre-CE systems have been proposed: when the secondary component is a planet or brown dwarf (Eggleton 1978; Livio & Soker 1984) or when the secondary is a low-mass red dwarf (Iben & Livio 1993). In both scenarios the primary is a red giant or supergiant, and we assume that during the CE phase some amount of mass is accreted onto the secondary.

In the first scenario (Livio & Soker 1984), the initial planet/brown dwarf mass of $\sim$ 0.0125 M$_\odot$ is greatly enhanced by accretion from the envelope of the giant until it becomes a low mass stellar companion to the red giant's core. If this scenario is correct, the small initial mass of the secondary implies that approximately 90 per cent of its final mass is accreted. Therefore, the isotopic ratios of carbon and oxygen would be expected to have values typical of red giant or supergiant stars, not red dwarfs.

The second scenario differs from the first since the secondary star is assumed to be a main-sequence star with a fully-convective (or thickly-convective) envelope. During the CE phase, we assume that the original surface of the red dwarf is heated by the overlying layers of accreted material. In this case, it has been shown that the star develops a highly inhomogeneous structure consisting of an interior formed by an almost unperturbed original star and a convective envelope composed of accreted matter (Prialnik & Livio 1985; Sarna & Ziółkowski 1988). These two regions are separated by a very stable (radiative) temperature inversion layer. Immediately after the CE phase, the secondary is out of thermal equilibrium due to the fact that the disturbance factors (accretion during the CE phase and/or heating by the subdwarf in planetary nebulae) still work and their timescale is very short ($10^3$–$10^4$ years) in comparison to the main-sequence thermal timescale ($t_{K-H} \sim 3 \times 10^7$ years for 1 M$_\odot$ – for lower mass main-sequence stars it is much longer). After a thermal timescale of the accreted envelope, the stable temperature inversion layer disappears and the secondary relaxes to its equilibrium radius. The material accreted during the CE phase will be mixed on a convective-mixing timescale (dynamical timescale). The region which is mixed consists of the convective region of the original star and the accreted matter.

The second scenario thus gives rise to two distinct observational cases, corresponding to the different evolutionary stages after CE phase. In the first case, when the secondary is out of thermal equilibrium, one would expect such systems to exhibit abundance ratios typical of red giants, not red dwarfs. These systems will appear as planetary nebulae with the secondary out of thermal equilibrium – the radius of the secondary star will be much greater than the radius expected for a main-sequence star in thermal equilibrium. The primary will be a hot subdwarf. In the second case, when the secondary has relaxed to thermal equilibrium, one would expect to observe abundance ratios intermediate between red dwarfs and red giants. These systems will appear as detached white-dwarf/red-dwarf binaries without planetary nebulae. The exact ratios will depend on the amount of mass accreted during the CE phase and the thickness of the convective zone in the pre-CE red dwarf.

For the modelling of this second scenario, we assume that the pre-CE red-dwarf has a mass (M$_{initial}$) in the range 0.08 to 0.7 M$_\odot$ and we assume initial red-dwarf carbon and oxygen isotopic ratios as for typical main-sequence stars. We further assume that the matter accreted by the red-dwarf comes from the red-giant's envelope, and therefore the accreted matter has isotopic ratios typical of a red-giant's envelope. The thickness of the convective zone is calculated using zero-age main-sequence models. In our models, a star with a mass of 0.3 M$_\odot$ is fully convective. The red dwarf models were computed using a standard stellar evolution code based on the Henyey-type code of Paczyński (1970), which has been adapted to low mass main-sequence stars.

Also, in our computations we have employed the mixing-length algorithm proposed by Paczyński (1969), using the opacity tables of Heubner et al. (1977). We assume Population I chemical composition for both components (X=0.7; Z=0.03) and for element abundances we adopt the Kurucz (1979) values for C and O: $X_C$=0.00397, $X_O$=0.00964. The results of our modelling are presented in figures 1 and 2, which show the abundance ratios for carbon and oxygen isotopes as a function of the initial ($M_{initial}$) and final ($M_{final}$) masses of the secondary star. If we determine the isotopic ratios $^{12}C/^{13}C$ and $^{16}O/^{17}O$ from infrared observations and we know the mass of the red dwarf in the PCEB, we can determine the red dwarf mass in the pre-CE system and the amount of mass accreted during the CE phase, but only for the second case when the red dwarf is in thermal equilibrium. In the first case, when secondary is a red dwarf out of thermal equilibrium, the accreted matter is not mixed with the interior of the star, therefore we can only say that some amount of mass has been accreted during the CE phase.

## 3 THE PREDICTION

Based on the above arguments and modelling, we make three predictions.

### 3.1 Red dwarfs in low-mass PCEBs

If Eggleton (1978) and Livio & Soker (1984) are correct, and the progenitor red dwarfs are planet/brown-dwarf type objects, then red dwarfs in low-mass PCEB's should exhibit abundance ratios of carbon and oxygen isotopes typical of red giant stars, since they have accreted a substantial amount of envelope material (the accreted mass constitutes maybe 90 per cent of the final secondary mass). The best candidates are AA Dor (Paczyński 1980), NN Ser (Catalán et al. 1994a) and RE1629+781 (Catalán et al. 1994b) because they have low-mass secondaries. Observationally these systems are probably too faint for measurements of the isotopic ratios to be made.

### 3.2 Red dwarfs in PCEBs with PNs

Red dwarfs in PCEBs with planetary nebulae, in which the secondary is not in thermal equilibrium, will also exhibit abundance ratios typical of red giants. The internal structure of the secondary star is inhomogeneous and therefore the accreted material cannot mix with the original star. Since the secondaries are out of thermal equilibrium, their radii are much larger than those typical of main-sequence stars with the same mass. Hence the best PCEB candidates are those that have over-large secondary radii and are surrounded by planetary nebulae, eg. V477 Lyr (Pollacco & Bell 1994), UU Sge (Pollacco & Hilditch 1994), BE UMa (Ferguson et al. 1987) and KV Vel (Landolt & Drilling 1986).

### 3.3 Red dwarfs in PCEBs without PNs

Red dwarfs in PCEBs without planetary nebulae, in which the secondary is in thermal equilibrium, should exhibit abundance ratios intermediate between red giants and solar values. The red dwarfs in these systems have radii typical of main-sequence stars and therefore the best candidate for this case is V471 Tau (eg. Young, Skumanich & Paylor 1988). Feige 24 (Paerels et al. 1986) and HZ 9 (Stauffer 1987) are also possible candidates, although their radii have not been determined accurately.

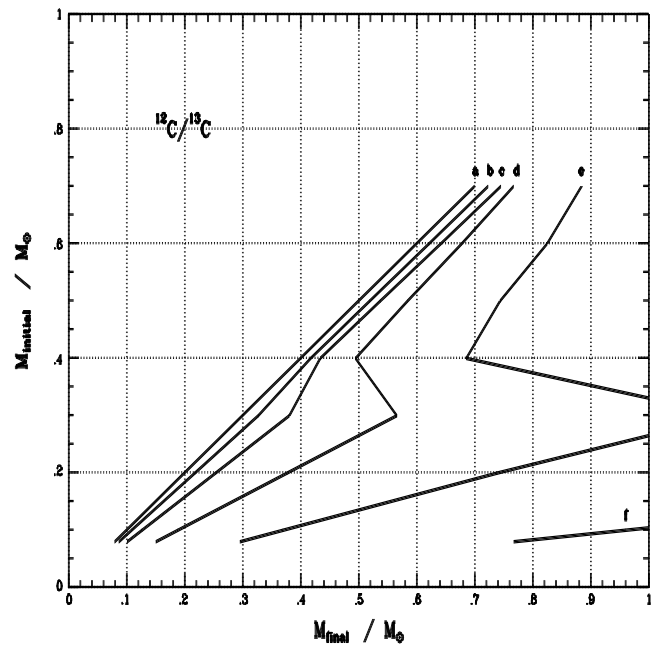

**Figure 1.** The isotopic ratio $^{12}C/^{13}C$ as a function of the pre-CE and PCEB red dwarf mass. The curves of constant isotopic ratio indicated by the letters a, b, c, d, e and f correspond to values of 91, 70, 50, 30, 20 and 16, respectively.

Using figures 1 and 2 in conjunction with isotopic ratios determined from observations, we will be able to determine the amount of mass accreted during the CE phase and also the pre-CE mass of the secondary. For example, if we assume that the secondary accretes an amount of mass comparable to its pre-CE mass, then in the case of V471 Tau (which has a mass of 0.73 $M_\odot$), $^{12}C/^{13}C$ and $^{16}O/^{17}O$ will have the approximate values 25 and 700, respectively, as opposed to 90 and 2724 (solar composition).

### 3.4 Infrared observations

The abundance ratios predicted above can be determined by infrared observations of CO bands. Specifically, the bands of $^{12}CO$ and $^{13}CO$ around 1.59$\mu$m, 2.3$\mu$m (2.29$\mu$m, 2.32$\mu$m, 2.35$\mu$m and 2.38$\mu$m) and 4.6$\mu$m (Harris and Lambert 1984ab; Bernat et al. 1979) are easily resolved by infrared echelle spectrographs (such as the Cooled Grating Spectrometer (CGS4) on the 3.8m United Kingdom Infrared Telescope (UKIRT) on Hawaii) and give a direct measurement of the $^{12}C/^{13}C$ ratio, and using stellar atmosphere models (comparing the observed spectrum to a synthetic one – see Harris & Lambert 1984ab) we can calculate the $^{16}O/^{17}O$ ratio.

Such an approach has already been attempted by Dhillon and Marsh (1994), but their detection of enhanced

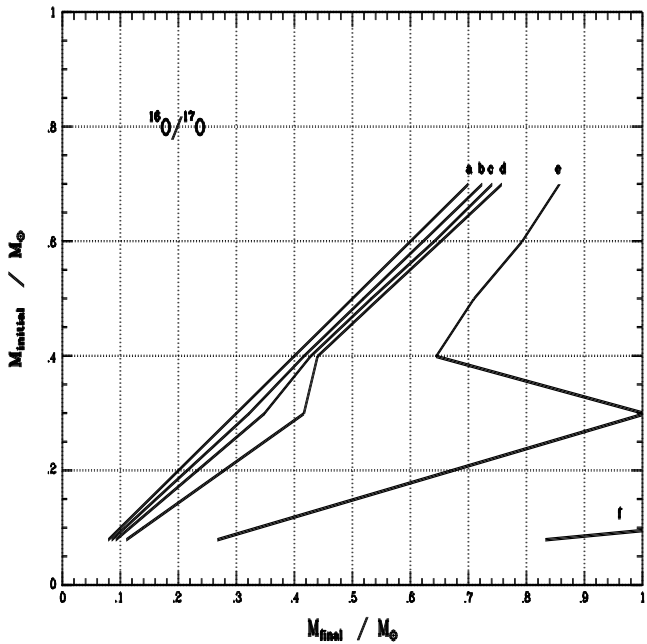

**Figure 2.** The isotopic ratio $^{16}O/^{17}O$ as a function of the pre-CE and PCEB red dwarf mass. The curves of constant isotopic ratio indicated by the letters a, b, c, d, e and f correspond to values of 2724, 2000, 1500, 1000, 500 and 400, respectively.

$^{13}$CO in the detached system V471 Tau is unfortunately inconclusive and requires further observations. As well as confirming the CE scenario, such observations will provide fundamental input to CE theories of CV evolution. For example, through the use of the computations presented in Figures 1 and 2 it will be possible determine the amount of mass accreted during the CE phase and the initial mass of the red dwarf component prior to the CE phase. It will also provide information on the structure and type of the red and white dwarfs in PCEB systems.

ACKNOWLEDGEMENTS

This work was supported in part by the Polish National Committee for Scientific Research under grant 2–2115–92–03. T. R. Marsh is supported by a PPARC Advanced Fellowship. P. B. Marks is in receipt of a PPARC studentship.